\documentclass[prd,aps,twocolumn
,preprintnumbers,nofootinbib,showpacs,amsmath,amssymb,superscriptaddress]{revtex4-1}
\usepackage{color}
\definecolor{purple}{rgb}{0.7,0.0,0.5}
\usepackage[bookmarks = false, colorlinks = true, linkcolor = blue, citecolor = purple, urlcolor=purple]{hyperref}

\usepackage{amsfonts}
\usepackage{amsmath}
\usepackage{amssymb}
\usepackage{upgreek}
\usepackage{bm}
\usepackage{dcolumn}
\usepackage{epsfig}
\usepackage{graphicx}
\usepackage{graphics}
\usepackage[latin1]{inputenc}
\usepackage{latexsym}
\usepackage{rotating}
\usepackage{hyperref}
\usepackage[all]{hypcap}
\usepackage{xspace}

\newcommand{\simgt}{\lower.5ex\hbox{$\; \buildrel > \over \sim \;$}}
\newcommand{\simlt}{\lower.5ex\hbox{$\; \buildrel < \over \sim \;$}}

\makeatletter
\AtBeginDocument{\@ifpackageloaded{natbib}{\ifNAT@numbers\if@filesw\immediate\write\@auxout{\string\global\string\NAT@numberstrue}\fi\fi}{}}
\makeatother
\begin{document}

\title{Constraint on modified dispersion relations for gravitational waves from
gravitational Cherenkov radiation
}

\author{Satoshi Kiyota}
\affiliation{
Department of Physical Sciences, Hiroshima University, 
 Kagamiyama 1-3-1, Higashi-hiroshima, 739-8526, Japan}
\author{Kazuhiro Yamamoto}
\affiliation{
Department of Physical Sciences, Hiroshima University, 
 Kagamiyama 1-3-1, Higashi-hiroshima, 739-8526, Japan}
\affiliation{
Hiroshima Astrophysical Science Center, Hiroshima University, 
 Kagamiyama 1-3-1,Higashi-hiroshima, 739-8526, Japan
}

\begin{abstract}
We investigate the hypothetical process of  gravitational Cherenkov 
radiation, which may occur in modified gravity theories. 
We obtain a useful constraint on a modified dispersion relation for 
propagating modes of gravitational waves, which could be predicted 
as a consequence of  violation of the Lorentz invariance in modified 
theories of gravity. 
The constraint from  gravitational Cherenkov radiation and that 
from direct measurements of the gravitational waves emitted by a 
compact binary system are complementary to each other. 
\end{abstract}

\maketitle

\def\bfp{{\bf p}}
Gravitational waves, predicted in  gravitational theories, are at the frontier of physics and 
astronomy.
Recently, a variety of the gravitational theories have been proposed, motivated by a possible 
explanation of the accelerated expansion of the universe. 
For example,  in recent research viable gravitational theories 
of  gravitational waves with a mass term have been proposed \cite{mg1,mg2,mg3}.   
Furthermore, graviton oscillations are predicted in the ghost-free bigravity theory
\cite{DeFelice,Narikawa}.
In the most general scalar tensor theory of the second derivative \cite{Horndesky}, 
which was rediscovered
recently \cite{Deffayet}, the propagation speed of  gravitational waves may deviate from the 
propagation speed of light  \cite{Kobayashi}.
Within general relativity, the propagation speed of  gravitational waves is 
the same as that of light, whose deviation is related to the breaking of  Lorentz 
invariance \cite{EinsteinAther1,EinsteinAther2}. 
Thus, the gravitational wave is important to characterize modified theories of gravity.

The dispersion relation for gravitational waves has been much argued in the literature. 
For example, constraints on the mass of  gravitational waves have been discussed \cite{Goldhaber}.
Although gravitational waves have not been directly detected,  the progress of observations
such as those from the advanced LIGO project and the KAGRA project will make direct measurements possible. 
Assuming such future prospects of observational experiments,  Mirshekari, Yunes, and 
Will investigated a future constraint on the modified dispersion relation 
through  gravitational wave experiments \cite{MirshkariYunesWill}. 
They demonstrated that a stringent constraint on the mass term can be obtained with 
future direct measurements of gravitational waves emitted by a compact binary system. 
The authors of the recent papers \cite{KYagi1,KYagi2} further investigated the 
orbital evolution of binary pulsars in modified gravity models with Lorentz symmetry 
breaking and obtained a useful constraint on the model parameters.

However, it is known that  gravitational Cherenkov radiation (GCR) arises if a 
massive particle moves faster than the speed of the gravitational waves, which is possible
when the propagation speed of the gravitational waves is smaller than that of light
\cite{Caves}. 
This hypothetical process puts a stringent constraint on the propagation speed 
of  gravitational waves by including  observations of  extremely high 
energy cosmic rays \cite{Moore}. 
The usefulness is indeed demonstrated for  modified gravity models of the
new Ether-Einstein gravity and the Galileon-type cosmological model \cite{Eliott,Kimura}. 

In the present letter, we investigate the constraint on the modified dispersion relation 
for gravitational waves from  GCR.
We assume the same modified dispersion relation for the gravitational waves in 
Ref.~\cite{MirshkariYunesWill}:
\begin{equation}
 \omega_k^2=  k^{2}c_s^{2}+m_{g}^{2}c_s^{4}+A{\bf}k^{\alpha}c_s^{\alpha},
\label{mdr}
\end{equation}
where $k$, $c_s$, and $m_g$ are the wave number, the propagation speed, and the 
mass of the gravitational wave, respectively, and $A$ and $\alpha$ are also the 
parameters of the modified dispersion relation. 
In the absence of the terms of $m_g$, $A$, and $\alpha$, it is known that  GCR puts the following
constraint on the propagation speed \cite{Moore}:
\begin{equation}
  1-c_{s}
  \hspace{0.3em}\raisebox{0.4ex}{$<$}\hspace{-0.75em}\raisebox{-.7ex}{$\sim$}\hspace{0.3em}
 2\times10^{-17}\biggl(\frac{10^{11}{\;\rm GeV}}{p}\biggr)^{3/2}\biggl(\frac{1\;{\rm Mpc}}{ct}\biggr)^{1/2}.
 \end{equation}  
However, the GCR process for the modified dispersion relation of the form
(\ref{mdr}) has not been discussed.
We demonstrate that  GCR puts a stringent constraint
on $A$ and $c_s$, depending on $\alpha$ and the sign of $A$, and that
GCR cannot put a useful constraint on $m_g$. From the 
direct measurement of  gravitational waves, the constraint on $m_g$ 
is stringent, but the constraint on $A$ is not very stringent \cite{MirshkariYunesWill}, 
which means that the two methods are complementary. 

\begin{figure}[t]
\begin{tabular}{c}
\begin{minipage}{0.5\hsize}
 \begin{center}
  \includegraphics[width=50mm]{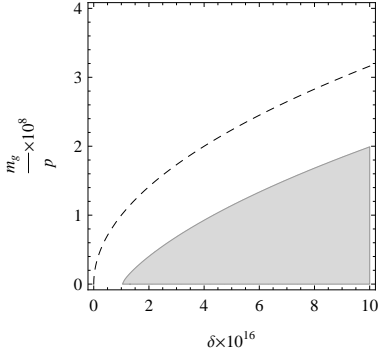}
 \end{center}
 \end{minipage}
\end{tabular}
\caption{
Constraint on the model parameters from  GCR for $m_g/p$ and 
$\delta$ in the case $A=0$. 
The shaded region satisfies (\ref{dEdtmg3}); this region is excluded 
by  GCR, where we assumed $p=10^{11}$~GeV and $ct=1$ Mpc.
The dashed curve shows  (\ref{GCRconditionmg1})
with $k=p$, above which the GCR process does not appear effectively. 
 \label{fig:mg_condition}
}
\end{figure}

Following Ref.~\cite{Moore,Kimura}, the rate of emitting energy through 
 GCR from a massive particle with mass 
$m$ and  relativistic momentum $p=|{\bf p}|$ can be written as
\begin{eqnarray}
&&{dE\over dt}=
G_Np^2\int_0^\infty dk k^2
\int_{-1}^{+1} d\cos\theta \sin^4\theta 
\nonumber\\
&&~~~~~~\times\delta_D(\Omega_i-\Omega_f-\omega_k),
\label{dEdt}
\end{eqnarray}
where $G_N$ is Newton's universal gravitational constant, 
$\Omega_i$ and $\Omega_f$
are defined by $\Omega_i=\sqrt{{\bf p}^2+m^2}$ and 
$\Omega_f=\sqrt{({\bfp}-{\bf k})^2+m^2}$, respectively, and 
$\delta_D$ is the Dirac delta function.
Using the identity $\delta_D(\Omega_i-\Omega_f-\omega_k)=~
2\Omega_f\delta_D(\Omega_f^2-(\Omega_i-\omega_k)^2)$,
we may write
\begin{eqnarray}
&&\delta_D(\Omega_i-\Omega_f-\omega_k)
={\Omega_f\over p k}\delta_D\biggl(\cos\theta-{k\over 2p}
\nonumber\\
&&
~~+{c_s^2k\over 2p}
\biggl(1+{c_s^2 m_g^2\over k^2}+A(c_sk)^{\alpha-2}\biggr)
\nonumber\\
&&
~~-{c_s\over p}\sqrt{p^2+m^2}
\sqrt{1+{c_s^2 m_g^2\over k^2}+A(c_sk)^{\alpha-2}}\biggr).
\end{eqnarray}
Integration of (\ref{dEdt}) makes a nontrivial contribution if
\begin{eqnarray}
&&
\cos\theta=\frac{k}{2p}-\frac{c_{s}^{2} k}{2p}\left(1+\frac{c^{2}_{s}m_{g}^{2}}{k^{2}}+A{\bf}(c_{s}k)^{\alpha-2}\right)
\nonumber\\
&&~~~+\frac{c_{s}}{p}\sqrt{\mathstrut p^{2}+m^{2}}\sqrt{\mathstrut 1+\frac{c^{2}_{s}m_{g}^{2}}{k^{2}}+A{\bf}(c_{s}k)^{\alpha-2}}\leq 1,
\label{cosc}
\end{eqnarray}
which is a condition at which  GCR arises.
Assuming $m/p\ll 1$, $m_g/k\ll 1$, $A(c_sk)^{\alpha-2}\ll1$, and $|\delta|\ll1$, 
where we defined $\delta=1-c_s^2$, we have
\begin{eqnarray}
\cos\theta\simeq 1+{(k-p)\over 2p}\left(\delta-{m_g^2\over k^2}-Ak^{\alpha-2}\right).
\label{cosa}
\end{eqnarray}
We may assume $k-p\leq0$, 
and then condition (\ref{cosc}) leads to
\begin{align}
\delta-{m_g^2\over k^2}-Ak^{\alpha-2}\geq0.
\label{GCRconditionmg}
\end{align}

First we consider the case $A=0$ but with finite graviton mass $m_g\neq0$ and 
$c_s<1$, i.e., $\delta>0$. In this case, (\ref{GCRconditionmg}) gives
\begin{align}
\delta-{m_g^2\over k^2}\geq0,
\label{GCRconditionmg1}
\end{align}
thus GCR does not appear when $\delta=0$.
We integrate Eq.~(\ref{dEdt}) by adopting the approximation
$\theta\ll1$, and we have
\begin{align}
\frac{dE}{dt}={G_{N}}\int_{m_g/\sqrt{\delta}}^{k_{\rm max}}dkk
\left(\left (k-p\right)\left(\delta-{m_g^2\over k^2}\right)\right)^{2},
\label{dEdtmg}
\end{align}
which yields 
\begin{eqnarray}
&&\frac{dE}{dt}={G_N\over 12}\biggl(27m_g^4-64\delta^{1/2}pm_g^3+36\delta p^2 m_g^2
\nonumber\\
&&~~~~~~~~+p^4\delta^2
-12m_g^2(m_g^2-2p^2\delta)\ln\left[{m_g\over p\delta^{1/2}}\right]\biggr),
\label{dEdtmg2}
\end{eqnarray}
where we set $k_{\rm max}=p$.
We estimate the condition at which the damping from  GCR is
significant for an extremely high energy cosmic ray with  initial 
energy $p$ during  time $t$ as ${dE/ dt}>  {p/ t}$,  
which yields
\begin{eqnarray}
&&
27m_g^4-64\delta^{1/2}pm_g^3+36\delta p^2 m_g^2+p^4\delta^2
\nonumber\\
&&~~
-12m_g^2(m_g^2-2p^2\delta)\ln\left[{m_g\over p\delta^{1/2}}\right]>   {12\over G_N}\frac{p}{t}.
\label{dEdtmg3}
\end{eqnarray}
The shaded region in Fig. \ref{fig:mg_condition} satisfies this condition (\ref{dEdtmg3}), 
which is excluded as the constraint of  GCR. 
Here we assume an extremely high energy cosmic ray of $p=10^{11}$~GeV 
from a cosmological distance $ct=1\;{\rm Mpc}$.
The dominant contribution of  the $k$ integration in Eq.~(\ref{dEdtmg})
comes from the region of $k$ of the order of $p$. 
The dashed curve in Fig. \ref{fig:mg_condition} shows the border of 
(\ref{GCRconditionmg1}) with  $k$ replaced by $p$, i.e., $\delta=m_g^2/p^2$. 
In the region above the dashed curve,  GCR does not occur effectively. 
Thus we obtain the constraint from GCR process for $\delta\simgt10^{-16}$; 
however, the constraint on  $m_g$ is of the order of $m_g \simlt 10^{3}$ GeV. 
When a high energy cosmic ray of $p=10^{3}$~GeV 
from a cosmological distance $ct=1{\rm Mpc}$ is assumed, 
we obtain the constraint on the $m_g$ is of the order of $m_g \simlt 10$ GeV
for $\delta\simgt10^{-4}$.
Thus the constraint on the graviton mass from  GCR is not very stringent.

Next we consider the case $m_g=0$; the condition at which  GCR arises, (\ref{GCRconditionmg}), is
\begin{align}
\delta-A{\bf}k^{\alpha-2}\geq 0. 
\label{GCRcondition}
\end{align}
One can observe that  GCR arises
even when $\delta<0,$ i.e., $c_s>1$, if $A$ is negative. 
This is quite a contrast to the usual conclusion that  GCR arises only 
when $c_s<1$ \cite{Caves,Moore}.
We integrate Eq.~(\ref{dEdt}),
adopting the approximation $\theta\ll1$, which gives
\begin{eqnarray}
\frac{dE}{dt}&=&{G_{N}}\int_{0}^{p}dkk\left(\left (k-p\right)\left(\delta-A{\bf}k^{\alpha-2}\right)\right)^{2},
\nonumber\\
&=&{G_{N}\over12}\biggl(\delta^{2}p^{4}+\frac{6A{\bf}^{2}p^{2\alpha}}{\alpha(\alpha-1)(2\alpha-1)}
\nonumber\\
&&
-\frac{48\delta A{\bf}p^{\alpha+2}}{\alpha(\alpha+1)(\alpha+2)}\biggr).
\label{dEdtalpha}
\end{eqnarray}
We here estimate the condition at which no significant radiation damping occurs 
from  GCR for an extremely high energy cosmic ray with initial 
energy $p$ during  time $t$ as
\begin{eqnarray}
\frac{dE}{dt} \leq
  \frac{p}{t}, 
\end{eqnarray}
which gives
\begin{eqnarray}
\delta^{2}p^{4}+\frac{6A{\bf}^{2}p^{2\alpha}}{\alpha(\alpha-1)(2\alpha-1)}
-\frac{48\delta A{\bf}p^{\alpha+2}}{\alpha(\alpha+1)(\alpha+2)}
\leq
  \frac{12p}{G_{N}t}.
\nonumber\\
\label{GCRcodition2}
\end{eqnarray}

\begin{figure}[t]
\begin{tabular}{c}
\begin{minipage}{0.5\hsize}
 \begin{center}
  \includegraphics[width=50mm]{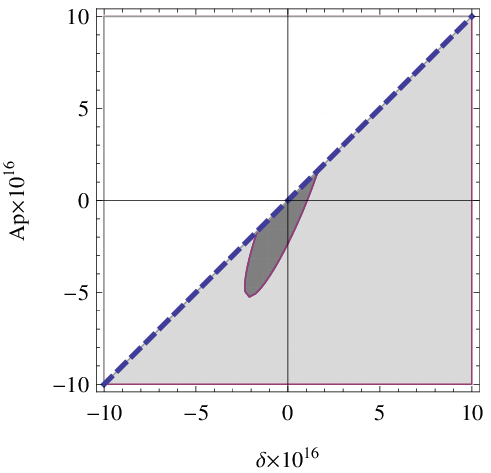}
 \end{center}
 \end{minipage}
\end{tabular}
\caption{Constraint on the model parameters from  GCR for the case $\alpha=3$ and $m_g=0$.
The oblique straight line is the border of (\ref{GCRcondition}) with  $k$ replaced by $p$, 
and  GCR does not occur in the upper region effectively. 
The elliptic curve is the border of (\ref{GCRcodition2}), 
inside of which (dark-gray region) satisfies (\ref{GCRcodition2}), where 
the GCR effect is so small that the extremely high energy cosmic rays 
do not damp. The light-gray region is excluded by  GCR.
\label{fig:Ap}
}
\vspace{3mm}
\begin{tabular}{c}
\begin{minipage}{0.5\hsize}
 \begin{center}
  \includegraphics[width=50mm]{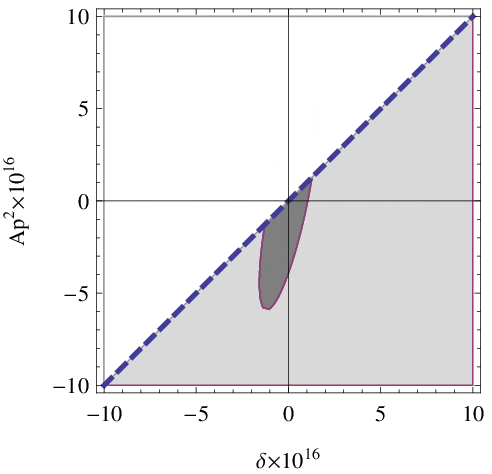}
 \end{center}
 \end{minipage}
\end{tabular}
\caption{Same as Fig.~\ref{fig:Ap}, but for the case $\alpha=4$.
\label{fig:App}
}
\end{figure}

We examine  conditions (\ref{GCRcondition}) and (\ref{GCRcodition2}). 
Here we assume that the extremely high energy cosmic rays of $p=10^{11}$~GeV 
from a cosmological distance $ct=1{\rm \;Mpc}$ are not significantly affected 
by the damping through  GCR \cite{Moore}.
Figures \ref{fig:Ap} and \ref{fig:App} show the constraint on the model parameters from  GCR
for  cases $\alpha=3$ and $4$, respectively. 
The oblique straight dashed line in each figure 
shows the border of  condition (\ref{GCRcondition})
with  $k$ replaced by $p$, above which  GCR does not arise effectively.
GCR arises in the lower region of this straight line.
The elliptic regions in Figs. \ref{fig:Ap} and \ref{fig:App} 
satisfy  condition  (\ref{GCRcodition2}), where  GCR arises but the effect is small. 
The lower light-gray region in each figure is excluded by the GCR process. 


For the case $\delta=0$, i.e., $c_s=1$, (\ref{GCRcodition2}) yields
\begin{eqnarray}
&&|A|
\leq 4\times
10^{5-11\alpha}\sqrt{\alpha(\alpha-1)(2\alpha-1)}
\nonumber\\
&&~~~~
\times\left(\frac{10^{11}\;{\rm GeV}}{p}\right)^{\alpha-1/2}
\left(\frac{1~{\rm Mpc}}{ct}\right)^{\frac{1}{2}}{\rm GeV}^{2-\alpha}.
\end{eqnarray}
Thus, for the case $c_s=1$, by combining  (\ref{GCRcondition}) and (\ref{GCRcodition2}),
the constraint from GCR excludes the value  
\begin{align}
Ap{\bf}\leq-2\times10^{-16}\left(\frac{10^{11}{\rm \;GeV}}{p}\right)^{3/2}
\left(\frac{1~{\rm Mpc}}{ct}\right)^{\frac{1}{2}}
\label{GCRC3}
\end{align}
for $\alpha=3$ and 
 \begin{align}
Ap^2{\bf}\leq -4\times10^{-16}\left(\frac{10^{11}{\rm \;GeV}}{p}\right)^{3/2}
\left(\frac{1~{\rm Mpc}}{ct}\right)^{\frac{1}{2}}
\label{GCRC4}
\end{align}
for $\alpha=4$, respectively. 

Note that  GCR excludes the negative region of the parameter of $A$
when $\delta=0$, i.e., $c_s=1$. When $\delta\neq0$, the constraint on 
$A$ depends on $\delta$. When $\delta>0$, a positive region of $A$ is 
excluded depending on the value of $\delta$.

Now we discuss the implication of our results for specific models with 
the modified dispersion relation. 
First, we consider the model in  broken-symmetry scenarios
\cite{Amelino,Amelino2}, in which 
Lorentz invariance is broken associated with the Planck energy scale:
\begin{align}
\omega_k^{2}= k^{2}+m_g^{2}+\eta\frac{\omega_k}{E_{p}}k^{2},
\end{align}
where $E_{p}$ is the Planck energy scale and $\eta$ is a nondimensional parameter.
This model corresponds to our model parameters  $\alpha=3$ and $A=\eta/E_p$ and 
$\delta=0$. From (\ref{GCRC3}), the following region is excluded:
\begin{align}
\eta\simlt-2\times10^{-8}\biggl({10^{11} {\rm \;GeV}\over p}\biggr)^{5/2}\biggl({1{\rm \;Mpc} \over ct}\biggr)^{1/2}.
\end{align}
The authors of Ref.~\cite{Amelino} argue that $\eta$ naturally takes a negative value 
in the broken-symmetry scenarios, and so  GCR puts a useful constraint on this scenario.
%
In the previous work in Ref.~\cite{Jacobson}, a constraint 
similar to ours  is obtained for the photon's modified 
dispersion relation, by comparing synchrotron radiation in the 
presence of the modified dispersion relation and  observations
from the Crab Nebula.

We next consider the scenario with an extra dimension
\cite{Sefiedgar}, which is an example of the case of $\alpha=4$:
\begin{align}
\omega_k^{2}=k^{2}+m_g^{2}-\alpha_{\rm edt}{\omega_k^{4}\over E_p^2}, 
\end{align}
where $\alpha_{\rm edt}$ is a nondimensional parameter. 
This model corresponds to our model parameters  $\alpha=4$ and 
$A=-\alpha_{\rm edt}/E_p^2$ and $\delta=0$. 
The constraint from  GCR (\ref{GCRC4}) excludes
the value
\begin{align}
\alpha_{\rm edt}
\simgt6\biggl({10^{11}{\rm \;GeV}\over p}\biggr)^{7/2}\biggl({1{\rm \;Mpc}\over ct}\biggr)^{1/2}.
\end{align}
GCR  constrains a positive value of $\alpha_{\rm edt}$;
otherwise,  GCR does not occur.
In this scenario of the modified dispersion relation, 
the sign of $\alpha_{\rm edt}$ is not specified.
The authors of Ref.~\cite{Sefiedgar} argue that the case $\alpha_{\rm edt}<0$ 
is relevant to  theories in which a generalized uncertainty principle is used. 
Therefore, unfortunately, our results cannot resolve this argument. 

As discussed in Ref.~\cite{MirshkariYunesWill}, 
there are other possible models that predict the modified dispersion relation.
For example, the Horava-Lifshitz gravity model predicts a modified dispersion 
relation \cite{Horava1,Horava2,Blas,Vacaru,Bogdanos}:
\begin{eqnarray}
\omega_k^2=c_s^2k^2+{\kappa^4k^4\over 16}\biggl(\mu\pm {2k\over\varpi^2}\biggr)^2-{\kappa^2\eta\over 2}k^6,
\end{eqnarray}
for the left-right polarization modes, where $\kappa$, $\mu$, $\varpi$, 
and $\eta$ are parameters.
The condition (\ref{GCRconditionmg})
at which GCR arises can be generalized as follows:
\begin{align}
\delta-{\kappa^4k^2\over 16}\biggl(\mu\pm {2k\over\varpi^2}\biggr)^2+{\kappa^2\eta\over 2}k^4\geq0.
\label{GCRconditionmg6}
\end{align}
Following the theory with the detailed-balance condition, $\eta=0$, 
the condition (\ref{GCRconditionmg6}) is not satisfied as long as 
$\delta\leq0$, which does not allow GCR process.
For the theory without the detailed-balance with $\eta>0$, the condition 
(\ref{GCRconditionmg6}) might be satisfied even when $\delta=0$.
For simplicity, we consider the case $\mu=0$, which reduces to an example of 
the case $\alpha=6$ with $A=\kappa^4/4\varpi^4-\kappa^2\eta/2$. 
From (\ref{GCRcodition2}), we have the constraint on $\delta$ and $A$. 
For the case $\delta=0$, GCR excludes the value
$A(=\kappa^4/4\varpi^4-\kappa^2\eta/2)<-7\times10^{-60}{\rm GeV}^{-4}$,
by adopting $p=10^{11}$~GeV and $ct=1$~Mpc.

The theory with noncommutative geometries, discussed in Ref.~\cite{Garattini1},
predicts a modified dispersion relation of the form (1) with $\alpha=4$ at the 
lowest order of perturbative expansion of the momentum. 
The sign of the parameter A is naturally predicted to be positive. 
In this case,  GCR does not appear as long as $c_s=1$.

In Ref.~\cite{MirshkariYunesWill}, future prospects of constraining
the modified dispersion relation from direct measurements of 
gravitational waves from binary systems with  the Advanced 
LIGO Project are addressed. With the method proposed, it is possible to put 
a stringent constraint on the graviton mass $m_g$. For example, 
for the case $\alpha=3$, the authors of Ref. \cite{MirshkariYunesWill} predict that the constraints at a level of 
$m_g<10^{-22}$--$10^{-25}$~eV will be obtained, depending on the target and 
the signal-to-noise ratio.
However, their constraint on $A$ is not stringent. The best constraint is
$|A|<10 ~{\rm GeV}^{-1}$ for the case $\alpha=3$.
However, the corresponding constraint on $A$ from  GCR is quite stringent, 
$A\simgt-10^{-27}~{\rm GeV}^{-1}$,
although the constraint on $m_g$ is not very stringent.
Thus, the two methods are complementary.

We have investigated the constraint on the modified dispersion relation for
gravitational waves from  GCR, assuming the form (\ref{mdr}).
The constraint on $m_g$ is not very tight; however, the constraint on $A$ 
from  GCR can be very stringent.
When the propagation speed is less than the velocity of light, $c_s<1$, 
 GCR may appear even when $A$ is positive, which is constrained through the GCR 
process.  
When $c_s=1$,  GCR puts a constraint on $A$ only when $A$ is negative.
This constraint is very stringent compared with that from 
direct measurements of  gravitational waves from binary systems as 
found in Ref.~\cite{MirshkariYunesWill} (see also \cite{KYagi1,KYagi2}).
Thus, the constraint from  GCR is complementary to that from 
 direct measurements of  gravitational waves. 

\vspace{1mm}
After  this paper was nearly completed, Ref. \cite{Kosteleky}, 
in which a similar problem  is investigated, appeared on the arXiv. 
We focused  on the modified
dispersion relation of the form (\ref{mdr}) but the authors of 
Ref. \cite{Kosteleky} investigate constraints on various Lorentz violation 
operators.

\vspace{1mm}
We thank R. Kimura, C. Yoo, and D. Blas for useful discussions and comments.
The research by KY is supported by a Grant-in-Aid for 
Scientific Research from the Japan Ministry of Education, Culture, 
Sports, Science and Technology (No. 15H05895). 

\end{document}